
\input phyzzx \pubtype={} \date={} \twelvepoint \font\bb=msym10
\def\tr{{\rm tr}} \def\dalemb#1#2{{\vbox{\hrule height .#2pt
        \hbox{\vrule width.#2pt height#1pt \kern#1pt
                \vrule width.#2pt}
        \hrule height.#2pt}}}
\def\square{\mathord{\dalemb{5.9}{6}\hbox{\hskip1pt}}}
\def\rn{Reissner-Nordstr{\" o}m} \def\bogo{Bogomol'nyi}
\def\fourth{{1\over 4}} \def\half{{1\over 2}} 
 \def\hnabla{\hat{\nabla}} 
\def\cF{{\cal F}} \def\cD{{\cal D}} \def\tr{{\rm tr}}

\REF\bogomolnyi{E.B. Bogomol'nyi, Sov. J. Nucl. Phys. {\bf 24} (1976)
449.} \REF\wo{E. Witten and D. Olive, Phys. Lett. {\bf 78B} (1978)
97.} \REF\sy{R. Shoen and S.T. Yau, Comm. Math. Phys. {\bf 65} (1979)
45; {\bf 79} (1981) 231.} \REF\witten{E. Witten, Comm. Math. Phys.
{\bf 80} (1981) 381.} \REF\motivation{C. Teitelboim, Phys. Lett. {\bf
69B} (1970) 240; S. Deser and C. Teitelboim, Phys. Rev. Lett. {\bf
39} (1977) 249; M. Grisaru, Phys. Lett. {\bf 73B} (1978) 249.}
\REF\sugrpos{G. Horowitz and A. Strominger, Phys. Rev. {\bf D27}
(1983) 2793; C.M. Hull, Comm. Math. Phys. {\bf 90} (1983) 545; S.
Deser, Phys. Rev. {\bf D27} (1983) 2805; C. Teitelboim, Phys. Rev.
{D29} (1984) 2763.} \REF\teitelboim{C. Teitelboim, Phys. Lett. {\bf
60B} (1977) 240.} \REF\chs{See {\it e.g.} Callan, Harvey and
Strominger, hep-th/9112030, in {\it String Theory and Quantum Gravity
`91}, Trieste (1991).} \REF\gh{G.W. Gibbons and C.M. Hull, Phys.
Lett. {\bf 109B} (1982) 190.} \REF\ghhp{G.W. Gibbons, G. Horowitz,
S.W. Hawking and M. Perry, Comm. Math. Phys. {\bf 88} (1983) 295.}
\REF\tod{K. Tod, Phys. Lett. {\bf 121B} (1983) 241.} \REF\cremm{E.
Cremmer, in {\it Superspace and Supergravity}, eds. S.W. Hawking and
M. Ro{\v c}ek (C.U.P. 1981).} \REF\gst{M. G{\" u}naydin, G. Sierra
and P.K. Townsend, Nucl. Phys. {\bf B242} (1984) 244; {\it ibid} {\bf
B253} (1985) 573; Phys. Lett. {\bf 133B} (1983) 72; {\it ibid} {\bf
144B} (1984) 41; Phys. Rev. Lett. {\bf 53} (1985) 322; Class. Quantum
Grav. {\bf 3} (1986) 763.} \REF\sierra{ G. Sierra, Phys. Lett. {\bf
157B} (1985) 379.} \REF\nest{J. Nester, Phys. Lett. {\bf 83A} (1981)
241.} \REF\myers{F.R. Tangerhlini, Nuovo. Cim. {\bf 27} (1963) 636;
R.C. Myers and M.J. Perry, Ann. Phys. (N.Y) {\bf 172} (1986) 304;
R.C. Myers, Phys. Rev. D{\bf 35} (1987) 455.} \REF\gibbo{G.W.
Gibbons, in {\it Supersymmetry, Supergravity and Related Topics},
eds. F. del Aguila, J. A. de Azc{\' a}rraga and L.E. Iba$\tilde{\rm
n}$ez, (World Scientific 1985).}   \REF\shiraishi{G.W. Gibbons, Nucl.
Phys. {\bf B207} (1982) 337.} \REF\gibtow {G.W. Gibbons and P.K.
Townsend, {\it Vacuum interpolation in supergravity via super
p-branes}, Phys. Rev. Lett. {\it in press}.}  \REF\lnw{K. Lee, V.P.
Nair and E.J. Weinberg, Phys. Rev. Lett. {\bf 68} (1992) 1100.}
\REF\perry{P. van Nieuwenhuizen, D. Wilkinson, and M.J. Perry, Phys.
Rev. {\bf D13} (1976) 778.} \REF\kt{D. Kastor and J. Traschen, Phys.
Rev. {\bf D46} (1992) 5399.}

\Pubnum={DAMTP R/93/27\cr UMHEP-371\cr NSF-ITP-93-125\cr
hep-th/9310118}

\titlepage \vskip -2em \title{Supersymmetric Self-Gravitating
Solitons}

\centerline{G.W. Gibbons${}^1$\foot{gwg1@amtp.cam.ac.uk }}

\centerline{D. Kastor${}^{2}$\foot{kastor@phast.umass.edu}}

\centerline{L.A.J. London${}^{1}$\foot{lajl1@amtp.cam.ac.uk}}

\centerline{P.K. Townsend${}^{1}$\foot{pkt10@phx.cam.ac.uk}}

\centerline{J. Traschen${}^{2}$\foot{lboo@phast.umass.edu}}

\address{${}^1$D.A.M.T.P., Silver Street\break
         University of Cambridge\break
         England}

\address{${}^2$Department of Physics and Astronomy\break
               University of Massachusetts\break
        Amherst, MA 01003}

\vfil

\abstract

We show that the `instantonic' soliton of five-dimensional Yang-Mills
theory and the closely related BPS monopole of four-dimensional
Yang-Mills/Higgs theory continue to be exact static, and stable,
solutions of these field theories even after the inclusion of
gravitational, electromagnetic and, in the four-dimensional case,
dilatonic interactions, provided that certain non-minimal
interactions are included. With the inclusion of these interactions,
which would be required by supersymmetry, these exact
self-gravitating solitons saturate a gravitational version of the
Bogomol'nyi bound on the energy of an arbitrary field configuration.

\endpage

\chapter{ Introduction}

Many non-linear field theories in flat spacetime have soliton
solutions (by which we mean localised solutions of finite energy).
Two examples, which will serve as a basis for our discussion are the
BPS monopoles of four-dimensional ($d=4$) Yang-Mills(YM)/Higgs theory
and instantons viewed as solitons of five-dimensional ($d=5$) YM
theory\foot{It is convenient to consider these examples together
because the $d=4$ Bogomol'nyi YM /Higgs equations are the
dimensionally reduced $d=5$ YM self-duality equations.}.  An
important feature in each of these systems is the existence of a
Bogomol'nyi bound [\bogomolnyi] on the energy in terms of the
topological charge of the configuration (monopole charge or instanton
number, respectively). Static soliton configurations saturate these
bounds. Mathematically this implies that these solutions satisfy
certain first-order equations (Bogomol'nyi or self duality,
respectively). One can find analytic solutions of these first-order
equations describing  single solitons, as well as static
multi-soliton configurations. These multi-soliton solutions are
possible because the solitons satisfy a force balance with respect to
each other.  For example, in the case of BPS monopoles, the repulsive
force due to the gauge field is exactly compensated by an attractive
force due to the scalar field.

One can ask whether a similar situation holds for self-gravitating
solitons. In general the answer is clearly no because coupling
gravity to the systems described above disturbs the force balance;
there will be a net attractive force between solitons and static
multi-soliton solutions will no longer exist.  We shall see below,
however, that the force balance can be restored and energy bounds
shown to exist, if one couples solitonic matter to the bosonic fields
of ($N=2$) {\it supergravity}.  Such a coupling is actually quite
natural. Witten and Olive [\wo] showed that, for the BPS monopole,
the \bogo\ bound may be derived by embedding the theory in one with
$N=2$ supersymmetry. The key element to this result is that the
topological charge of the soliton appears as a central charge in the
supersymmetry algebra. We might expect that coupling this globally
supersymmetric matter theory to $N=2$ supergravity will lead to
energy bounds for gravitating BPS monopoles.

In general relativity, the energy and momentum of an asymptotically
flat four dimensional spacetime are given by the ADM $4$-momentum,
which is a surface integral over the $2$-sphere at spatial infinity.
An analogue of a \bogo\ bound for a gravitating soliton should be a
variant of the positive energy theorem [\sy,\witten], which bounds
the ADM $4$-momentum in terms of the topological charge of the
solitonic matter fields. Coupling to supergravity is natural in this
context as well.  Witten's [\witten] proof of the positive energy
theorem was motivated by [\motivation] and is most clearly understood
in the context of supergravity [\sugrpos].  In supergravity, the
global supercharges are also given by surface integrals at spatial
infinity [\teitelboim]. The supercharges combine with the ADM
$4$-momentum to form the global supersymmetry algebra.  We expect
that, as in the case of global supersymmetry [\wo], the topological
charge of the matter fields will enter this algebra as a central
charge.

The $N=2$ supergravity multiplet in four dimensions contains the
graviton, a pair of gravitini and a $U(1)$ gauge field, which we will
refer to as the Maxwell field\foot{We will be considering ungauged
supergravity in which the gravitini are neutral with respect to the
Maxwell field.  Coupling the gravitini to the Maxwell field
introduces a cosmological constant.}. The Maxwell field introduces
another force, which can restore the force balance between
solitons\foot{We note that over the past few years many gravitating
soliton solutions have been found in supergravity theories arising in
the low energy limit of string theory [\chs].  In many of these, the
force balance is restored by the antisymmetric tensor field.}. If we
restrict our attention to purely bosonic configurations then we are
dealing with solitonic matter coupled to Einstein/Maxwell theory.
Gibbons and Hull [\gh] have proven a version of the positive energy
theorem for Einstein/Maxwell theory, which we here generalise to
include a dilaton with arbitrary coupling constant $b$ (as defined by
the action $(6.1)$ to follow). Given certain conditions on the
stress-energy and charge current of the matter  fields, one has that
$$ M\ge {1\over \sqrt{1+b^2}}\sqrt{Q^2 +P^2}\ , \eqn\bound $$ where
$M$ is the ADM mass and $Q$ and $P$ are the total electric and
magnetic charges of the spacetime with respect to the Maxwell field.
In the context of supergravity, $Q$ and $P$ enter the algebra of
global supersymmetry transformations as central charges. Spacetimes
which saturate the bound \bound\ are those which have spinor fields
which are constant with respect to a `supercovariant' derivative
operator. This property can be thought of as an analogue of the first
order \bogo\  equations. The puzzle remains of how the topological
charge of the matter becomes a source for the Maxwell  field. As we
shall see below, a certain non-minimal Maxwell/matter interaction
achieves precisely this. This interaction might ordinarily be
disregarded because it is non-renormalisable, but it is required by
$N=2$ local supersymmetry. Recall that the Maxwell field is in the
same supermultiplet as the graviton so, in the context of
supersymmetry, it is not surprising that it should have
non-renormalisable interactions with matter fields.

In Section 2, we look at the case of $SU(2)$ YM/Higgs matter fields
in $4$ dimensions, coupled to gravity and an additional Maxwell field
but {\it without} a dilaton. In this case, we are able to show that
an energy bound on solitons exists, but we are unable to find
analytic solutions saturating this bound. In an appendix, we present
a proof that such solutions do exist, at least for sufficiently weak
values of the gravitational coupling. In Sections 3, 4 and 5, we
establish results for instantonic solitons in $d=5$ supergravity/YM
theory. In this case we are able to find analytic solutions for the
matter configurations. The solutions we find are non-singular,
static, self-gravitating, particle-like configurations with a (multi)
instanton core. For arbitrary core radius the solutions have the
property that they saturate a Bogomol'nyi - type bound, which we
derive following the method used for $d=4$ Einstein/Maxwell theory in
[\gh]. In the context of $d=5$ supergravity the solutions preserve
half the supersymmetry of the vacuum. For this reason we refer to
configurations which saturate the bound as `supersymmetric'.

We then return in Sections 5 and 6 to the $d=4$ case, but now
including a dilaton field. We first derive the bound \bound\ and find
the vacuum solutions which saturate this bound. These are previously
constructed charged dilaton black holes which reduce to the standard
\rn\ (RN) solution in the $b\rightarrow 0$ limit. The inclusion of
solitonic matter in the form of YM/Higgs monopoles is consistent with
the continued existence of a super-covariantly constant spinor
provided that the dilaton couples to the YM/Higgs fields in a
particular way. A surprising feature is that, once the dilaton/matter
coupling is fixed in this way, the $b\rightarrow 0$ limit cannot be
taken because the action would be singular. The results with and
without dilaton are therefore qualitatively different and must
therefore be considered separately. The case without a dilaton will
be considered in Section 2, as mentioned above. In Section 7  we
analyse the dilatonic case. In both cases self-gravitating monopoles,
should they exist, saturate a gravitational analogue of the
Bogomol'nyi bound, but we are able to find exact solutions of the
curved space YM/Higgs equations that saturate this bound only in the
dilatonic case. In fact, in the dilatonic case one finds that these
equations are solved by solutions of the {\it flat space} Bogomol'nyi
equations. (Multi) monopole solutions of the latter are well known
and the existence of self-gravitating (multi) monopoles saturating
the gravitational version of the Bogomol'nyi bound are therefore
guaranteed. The resulting $4$-metrics are non-singular and without
horizons. Without a dilaton, as shown in Section 2, the curved space
YM/Higgs equations are still solved by any solution of a version of
the Bogomol'nyi equations, but the latter are no longer the flat
space equations and their solution requires the simultaneous solution
of the Einstein equations. The fact that {\it exact} solutions of the
curved space YM/Higgs equations can be found in the dilatonic case
can be seen to be a consequence, at least for a particular value of
$b$, of our $d=5$ results. This is because certain results for $d=4$
can be obtained from $d=5$ by dimensional reduction; one simply
solves the $d=5$ equations for spaces of topology {\bb R}${}^3\times
S^1$ instead of {\bb R}${}^4$. The existence of exact
self-gravitating (multi) monopole solutions for a particular $d=4$
theory is therefore guaranteed from the $d=5$ results. Our results
obtained directly in $d=4$ are consistent with this observation.

\chapter{Four Dimensional Supergravitating Monopoles}

In this section, we will search for self-gravitating monopole
solutions in $3+1$ dimensions which saturate an energy bound. The
strategy we follow here will be repeated in other contexts in later
sections. In the present case, however, the calculations necessary to
establish the new results will be shorter than those in subsequent
sections because we can take advantage of a number of already
established results. This section can then serve the reader as a less
technical introduction to some of the methods used in the remainder
of the paper.

In $3+1$ dimensions, the positive energy theorem establishes that the
norm of the ADM four momentum $P_\mu$ satisfies $P_\mu P^\mu\le 0$
(with signature ($-+++$)), with equality only for Minkowski space. A
stronger result can be derived for Einstein/Maxwell theory
[\gh,\ghhp]. If $Q$ is the total electric charge of the Maxwell
field\foot{We assume zero magnetic charge for simplicity} and
$M^2\equiv -P_\mu P^\mu$, then we have  $$  M\ge \mid Q\mid\ .
\eqn\chargebound $$ We are looking for `ground states', i.e.
configurations which minimise the ADM mass subject to fixed electric
charge. The bound \chargebound\ is known to be saturated if and only
if the spacetime admits a complex spinor field $\epsilon$ which is
constant with respect to the supercovariant derivative  $$
\hnabla_\mu = \nabla_\mu + {i\over 4} F_{\alpha\beta}
\Gamma^\alpha\Gamma^\beta\Gamma_\mu\  , \eqn\supercov $$ where
$F_{\alpha\beta}$ is the Maxwell field strength.  One class of such
spacetimes are the extremal \rn\ solutions and their multi-black hole
generalisations the, Majumdar-Papapetrou solutions.  These are
electro-vac solutions. In this paper, however, we are interested in
spacetimes containing solitonic matter fields and for which the bound
\chargebound\ is still saturated.

In $3+1$ dimensions, the form of the most general metric and gauge
field admitting a supercovariantly constant spinor has been given in
[\gh,\tod].  If we restrict our attention to static metrics and gauge
fields, then these have the form $$ ds^2 = - e^{-2\phi}dt^2
+e^{2\phi}d{\bf x}\cdot d{\bf x}\qquad A=\pm e^{-\phi} dt\ , \eqn\mpl
$$ where $\phi =\phi\left( {\bf x}\right)$ and $d{\bf x}\cdot d{\bf
x}$ is the Euclidean $3$-metric. We can now find what matter sources
are required to produce fields of this form by simply demanding that
the Einstein-Maxwell equations be solved and then determining the
sources. From the Einstein and Maxwell constraint equations, we then
find that the charge density $\rho_q$ and energy density $\rho_m$ of
the matter fields (excluding the energy density of the Maxwell field)
are equal (up to a sign) and given by $$ \mid\rho_q\mid = \rho_m=
-{1\over 4\pi G}e^{-3\phi}\nabla^2\left( e^\phi\right)\ , \eqn\poisson
$$ where $\nabla^2$ is the Laplacian for Euclidean $3$-space.
Furthermore, one finds that the Einstein and Maxwell evolution
equations imply that the spatial components of the matter stress
energy $T_{ij}$ and the charge current $J_i$ all vanish $$
T_{ij}=J_i=0\ . \eqn\einstein $$ These conditions on the matter
sources characterise charge equal mass (or $q=m$) dust. Moreover, any
spatial configuration of $q=m$ dust solves the Einstein-Maxwell-matter
system and saturates the energy bound. One can simply choose the dust
configuration by specifying a function $f\left( {\bf x}\right)$,
solve $\nabla ^2\left( e^\phi\right)=-4\pi G f$, to obtain the metric
function, and recover the energy and charge densities via $\rho_m=|
\rho_q| = e^{-3\phi}f $. Note that $f\left({\bf x}\right)=0$ yields
the equation $\nabla^2 \left( e^\phi\right) =0$, the solutions of
which are the Majumdar-Papapetrou electro-vac spacetimes.

Next we shall show that solitonic matter fields saturating an
appropriate \bogo\ constraint can have energy and charge densities of
the $q=m$ dust form. We consider YM/Higgs matter coupled to
Einstein/Maxwell theory via the action $$ \eqalign{ S&={1\over 4\pi
G}\int\! d^4x\, e\left\{\fourth R-\fourth
F_{\mu\nu}F^{\mu\nu}\right\} \cr &+{1\over g^2}\int\!
d^4x\,\left\{-\fourth e\,\tr \left( G_{\mu\nu}G^{\mu\nu}\right)
-\half e\,\tr\left( D_{\mu}\Phi D^{\mu}\Phi\right) +{1\over
4}\epsilon^{\mu\nu\alpha\beta}F_{\mu\nu}\tr\left(\Phi
G_{\alpha\beta}\right)\right\}\ ,} \eqn\actionx $$ where
$e=\sqrt{-\det g_{\mu\nu}}$ with $g_{\mu\nu}$ the spacetime metric,
$G_{\mu\nu}$ is the covariant field strength for a Lie-algebra-valued
YM vector potential $B_\mu$ $$
G_{\mu\nu}=\partial_{\mu}B_{\nu}-\partial_{\nu}
B_{\mu}+[B_{\mu},B_{\nu}]\ , \eqn\geqf $$ $\Phi$ is the Higgs field
in the adjoint representation and $D_\mu\Phi$ its YM covariant
derivative $$ D_{\mu}\Phi=\partial_{\mu}\Phi +[ B_{\mu},\Phi ]\ .
\eqn\deqf $$ The YM and Higgs field are assumed to have dimensions of
$L^{-1}$, so the $d=4$ YM coupling constant $g$ has dimensions of
$(ML)^{-(1/2)}$. The coupling between the Maxwell and the YM/Higgs
fields is motivated by supersymmetry; in a supersymmetric theory the
coefficient of this term is non-zero and our results lead us to
believe that the value appearing in \actionx\ is that singled out by
supersymmetry\foot{This could be verified by a construction of the
general $N=2$ supergravity/YM/Higgs theory in the conventions of this
paper.}. It is helpful to keep in mind that the Coulomb force law for
the Maxwell field normalized as in \actionx\ is $F=GQ_1Q_2/r^2$.

Observe, that to preserve parity we require either $\Phi$ or $A_\mu$
to be parity odd. Since the $N=2$ Maxwell supermultiplet contains
both a scalar and a pseudoscalar Higgs field either one can be used
in a monopole solution. If we use the scalar then we should interpret
$A_\mu$ as the magnetic vector potential rather than the electric
vector potential. We shall suppose here that $\Phi$ is a pseudoscalar
and $A_\mu$ the electric potential.

Consider static configurations of the YM and Higgs fields, having $$
D_0 \Phi=0,\qquad G_{0i}=0\ , \eqn\static $$ and take the metric and
Maxwell field to have the form \mpl . The charge density $\rho_q$,
found by varying the action with respect to the Maxwell field is
given by $$ 4\pi G\rho_q = -{1\over2g^2}e^{-3\phi}\varepsilon^{ijk}
\tr\left( D_i\Phi G_{jk}\right)\ , \eqn\cdensity $$ where
$\varepsilon^{ijk}$ is the $3$-dimensional alternating tensor
density. The matter energy density $\rho_m$ is given by  $$ \eqalign{
4\pi G\rho_m &=\half\tr\left( D_i\Phi D^i\Phi\right)
+\fourth\tr\left( G_{ij}G^{ij}\right)\cr &={1\over
2g^2}e^{-2\phi}\tr\left[ D_i\Phi D_i\Phi +{1\over 2}e^{-2\phi}
G_{ij}G_{ij}\right]\ ,} \eqn\edensity $$ where the latter equality is
found by making use of the fact that $g_{ij}=\delta_{ij}e^{2\phi}$.
Now, if the YM and Higgs field satisfy a curved space form of the
\bogo\ equation, specifically $$
\sqrt{g^{{}_{(3)}}}g^{il}g^{jm}G_{lm}=\pm\varepsilon^{ijk}D_k\Phi\ ,
\eqn\cbogo $$ where $g^{{}_{(3)}}$ is the determinant of the
$3$-metric $g_{ij}$, then the expression for the energy density
\edensity\ becomes identical to that for the charge density
\cdensity\ and we have $\rho_m=\rho_q$. One can also check that,
given \cbogo , the spatial components of the matter stress energy
$T_{ij}$ and charge current $J_i$ vanish. Note that since
$g_{ij}=e^{2\phi}\delta_{ij}$, and $\varepsilon^{ijk}$ is independent
of the metric, \cbogo\ is equivalent to the Euclidean $3$-space
equation  $$ G_{ij}=\pm e^{\phi}\varepsilon^{ijk}D_k\Phi\ .
\eqn\anpbogo $$

In addition, given the form of the metric and the Maxwell field, then
if \cbogo\ is satisfied, one can check that both the YM and Higgs
equations of motion are satisfied as well. Unfortunately, the curved
space \bogo\ equation \anpbogo\ depends on the metric function
$e^\phi$, and we have not been able to find analytic solutions to
this coupled system.  However, specialising to the case of spherical
symmetry, we can show that solutions exist, at least for sufficiently
weak gravitational coupling. We give a proof of this in the Appendix.
We shall show later that when a dilaton field is included, the curved
space \bogo\ equation becomes identical to the flat space one, and so
exact solutions can easily be found.

We end this section by presenting an alternative derivation of the
results presented above that is closer in spirit to Bogomol'nyi's
original argument [\bogomolnyi]. We insert the result \mpl\ into the
matter part of the action \actionx. From the corresponding
Hamiltonian the total energy is  then found to be   $$  \eqalign{ E=
{1\over g^2}\int &d^3x\bigg\{ {1\over4} e^{-2\phi} \tr\big( G_{ij}\mp
e^{\phi}\varepsilon^{ijk}D_k\Phi\big)^2\cr
&+{1\over2}e^{2\phi}\big[\tr (G_{0i} G_{0i}) + e^{2\phi}\tr (D_0\Phi
D_0\Phi)\big] \pm\eta \partial_i\left[ e^{-\phi} B_i^{mat}\right]
\bigg\}\ ,} \eqn\hamx $$ where $$ B_i^{mat}={1\over
2\eta}\varepsilon^{ijk}\tr\left(\Phi G_{jk}\right)\ , \eqn\fgtr $$ is
the `matter' magnetic field. Assuming that ${\bf B}^{mat}\sim 1/r$
and that $\phi\rightarrow 0$ as $r\rightarrow \infty$, we immediately
derive the bound $$ E\ge {\eta\over g^2}\mid P^{mat}\mid\ ,
\eqn\hboundd $$ where $P^{mat}$ is the (dimensionless) total `matter'
magnetic charge determined by the flux of ${\bf B}^{mat}$ through the
$2$-sphere at spatial infinity $S_2$ $$ P^{mat}=\int_{S_2}\!
B_i^{mat}dS_i\ . \eqn\jku $$ By integrating \cdensity\ we see that $$
Q={\eta\over g^2}P^{mat}\ , \eqn\jklo $$ and so we have $E\ge Q$, in
agreement with \chargebound . The bound \hboundd\ is saturated when $$
G_{0i}=0, \qquad D_0\Phi=0,\qquad G_{ij}=\pm
e^{\phi}\varepsilon^{ijk}D_k\Phi\ , \eqn\eqswd $$ in agreement with
\static\ and \anpbogo .

\chapter{An Energy Bound For d=5 Einstein/Maxwell Theory}

The bosonic sector of $d=5$ supergravity coupled to matter
[\cremm,\gst,\sierra] has the action $$ S={1\over 4\pi G_5}\int\!
d^5x\bigg\{ -{1\over4}eR -{1\over4}eF_{\mu\nu}F^{\mu\nu} - {1\over
6\sqrt{3}}\varepsilon^{\mu\alpha\beta\gamma\delta} A_\mu
F_{\alpha\beta}F_{\gamma\delta} \bigg\} \ +\ S_{matter}\ .
\eqn\actionf $$ Here, $R=g^{\mu\nu}R^\lambda{}_{\mu\lambda\nu}$ is
the scalar curvature, $F_{\mu\nu}=2\partial_{[\mu}A_{\nu]}$ is the
Maxwell field-strength tensor, and $e=\sqrt{-g}$ is the determinant
of the f{\" u}nfbein $e_\mu{}^{\underline\alpha}\, $, where an
underlined index refers to the locally inertial Lorentz frame. For
$d=5$ we shall use the `mostly minus' metric signature. Note that
$A_{\mu}$ is a dimensionless pseudovector. The dimensions of $G_5$
(the $d=5$ Newton's constant) are $L^2M^{-1}$.

The Einstein and Maxwell field equations are $$ \eqalign{ G_{\mu\nu}
-2T_{\mu\nu}(F) &= 8\pi G_5 T_{\mu\nu}(mat.)\cr \nabla_\mu F^{\mu\nu}
-{1\over2\sqrt{3}}e^{-1}
\varepsilon^{\nu\alpha\beta\gamma\delta}F_{\alpha\beta}
F_{\gamma\delta}&= 4\pi G_5 J^\nu(mat.),} \eqn\einf $$ where $$
T_{\mu\nu}(F) = -\left(F_{\mu\lambda}F_{\nu}{}^\lambda
-{1\over4}g_{\mu\nu} F^{\alpha\beta}F_{\alpha\beta}\right),
\eqn\fieldf $$ and $$ T_{\mu\nu}(mat.)\equiv {1\over 2e} {\delta
S_{matter}\over\delta g^{\mu\nu}} \qquad J^\mu (mat.)\equiv {1\over
e}{\delta S_{matter}\over \delta A_\mu }\ . \eqn\varnf $$

We are now going to establish a Bogomol'nyi-type bound on the energy
of any field configuration, following the derivation in [\gh] of a
similar bound for $d=4$ Einstein/Maxwell theory. To this end we
introduce the Nester-like tensor [\nest] $$ \hat E^{\mu\nu}
={1\over2}\bar\epsilon \Gamma^{\mu\nu\rho}\hat\nabla_\rho\epsilon \ +
c.c.\ , \eqn\nestf $$ where $\epsilon$ is a complex commuting
$SO(1,4)$ spinor, $\bar\epsilon$ its Dirac conjugate $\epsilon^{\dag}
\Gamma^{\underline{0}}$ and $$ \hat\nabla_\mu\epsilon \equiv
\nabla_\mu\epsilon +{1\over 4\sqrt{3}}\big(
\Gamma^{\alpha\beta}\Gamma_\mu + 2\Gamma^\alpha
\delta^\beta_\mu\big)\epsilon F_{\alpha\beta}\ . \eqn\covf $$ Here
$\Gamma^\alpha$ is a Dirac matrix and
$\Gamma^{\alpha_{{}_1}\cdots\alpha_n}$ is the antisymmetrised product
of $n$ Dirac matrices (with `strength one'). The product of all five
Dirac matrices equals the unit matrix up to a sign; we choose the
sign such that $e\Gamma^{\alpha_1\dots
\alpha_5}=\varepsilon^{\alpha_1\dots \alpha_5}$. The covariant
derivative $\nabla$ is defined in terms of the usual (torsion free)
spin connection $\omega$. The definition \covf\ is directly motivated
by $d=5$ supergravity since the supersymmetry transformation of the
(complex) gravitino in that theory, in a background for which the
gravitino vanishes, is just $\delta\psi_\mu = \hat\nabla_\mu\epsilon$
(although $\epsilon$ would be anticommuting in this context).

Let $\Sigma$ be a spacelike hypersurface with hypersurface element
$dS_\mu$ and with boundary $\partial\Sigma$ at spatial infinity. If
the spacetime contains black holes, then there will also be
components of $\partial\Sigma$ corresponding to the black hole
horizons.  As shown in reference [\ghhp], the boundary terms on the
horizons can be made to vanish if $\epsilon$ satisfies
$\Gamma^{{\underline n}}\Gamma^{\underline 0}\epsilon=\epsilon$ where
${\underline n}$ indicates the direction normal to the horizon.  We
shall assume that this condition holds below. With this in mind we
consider the integral $$ \int_\Sigma \!\! dS_\mu\, e\nabla_\nu\hat
E^{\mu\nu} = \int_\Sigma \!\! dS_\mu \partial_\nu\big(e\hat
E^{\mu\nu}\big) ={1\over2}\int_{\partial\Sigma}\!\!dS_{\mu\nu}\,
e\hat E^{\mu\nu}\ , \eqn\hypf $$ where $dS_{\mu\nu}$ is the element
of the $3$-sphere at spatial infinity. Now, $\hat E^{\mu\nu}$ can be
written as $$ \hat E^{\mu\nu}={1\over2}\left\{
\bar\epsilon\Gamma^{\mu\nu\rho}\nabla_\rho\epsilon
-{\sqrt{3}\over4}\bar\epsilon\Gamma^{\mu\nu\alpha\beta}\epsilon
F_{\alpha\beta} -{\sqrt{3}\over2}\bar\epsilon\epsilon
F^{\mu\nu}\right\}\ +c.c.\ . \eqn\enef $$ Assuming that the spacetime
is asymptotically flat and that, asymptotically, $F$ is purely
electric and $\epsilon$ approaches the {\it constant} spinor
$\epsilon_{{}_\infty}$, we have that $$ \eqalign{ \int_\Sigma \!\!
dS_\mu\, e\nabla_\nu\hat E^{\mu\nu} &=
\int_{\partial\Sigma}\!\!dS_{\mu\nu}\left[{1\over4} e\bar
\epsilon_{{}_\infty}\Gamma^{\mu\nu\rho}\nabla_\rho\epsilon_{{}_\infty}
\; +\, c.c.\right] -{\sqrt{3}\over
4}(\bar\epsilon_{{}_\infty}\epsilon_{{}_\infty})
\int_{\partial\Sigma}\!\!dS_{\mu\nu}\, e F^{\mu\nu}\cr
&=\left(\bar\epsilon_\infty\Gamma^{\mu}\epsilon_\infty\right)P_{\mu}^{ADM}
-{\sqrt{3}\over 2}\left(\bar\epsilon_\infty\epsilon_\infty\right)Q\
,} \eqn\dervf $$ where, in the second line, we have used the
Witten-Nester expression for the ADM $5$-momentum, and $Q={1\over
2}\int_{\partial\Sigma}\! dS_{\mu\nu}\left( eF^{\mu\nu} \right)$ is
the total electric charge.

We turn now to the evaluation of $\nabla_\nu\hat E^{\mu\nu}$. A
lengthy calculation yields the result $$ \nabla_\nu\hat
E^{\mu\nu}=\left\{{1\over2}(\overline{\hat\nabla_\nu\epsilon})\Gamma^{\mu\nu\rho}
\hat\nabla_\rho\epsilon + c.c.\right\} + (4\pi G_5)K^{\mu}\ ,
\eqn\longf $$ where we have used the Einstein/Maxwell field equations
\einf\ and $$ K^{\mu}=\bar\epsilon\Gamma^{\nu}\epsilon
T_{\nu}{}^{\mu}(mat.)+{\sqrt{3}\over 2} \bar\epsilon\epsilon
J^{\mu}(mat.)\ . \eqn\keqnm $$ Combining \longf\ with \dervf\ we
deduce that $$ \eqalign{
\left(\bar\epsilon_\infty\Gamma^{\mu}\epsilon_\infty\right)P_{\mu}^{ADM}
-{\sqrt{3}\over 2}\left(\bar\epsilon_\infty\epsilon_\infty\right)Q
&=\int_{\Sigma}\! dS_{\mu}e\left[\half\overline{
\left(\hat\nabla_{\nu}\epsilon\right)}
\Gamma^{\mu\nu\rho}\left(\hat\nabla_{\rho}\epsilon\right)\
+c.c.\right]\cr &+\left( 4\pi G_5\right)\int_{\Sigma}\!
dS_{\mu}eK^{\mu}\ .} \eqn\keqn $$ The first term on the right hand
side of \keqn\ is non-negative for spinors $\epsilon$ satisfying the
(modified) Witten condition [\witten] $$ n\cdot\hat\nabla\epsilon =0\
, \eqn\lcond $$ where $n$ is a $5$-vector normal to $\Sigma$; it
vanishes if and only if $$ \hat\nabla_{\mu}\epsilon =0\ . \eqn\kill $$
We shall call a spinor satisfying \kill\ a Killing spinor.

The second term on the right hand side of \keqn\ is non-negative if
$K^\mu$ is future-directed timelike, or zero, for all $\epsilon$, and
we shall assume this condition in what follows. This term vanishes if
and only if $K^\mu =0$\foot{Choose a frame for which $K^i=0$. Then,
since $K^0\ge 0$, the integral $\int_{\Sigma}\! dS_{\mu}eK^{\mu}$
vanishes if and only if $K^0\equiv 0$. But if the integral vanishes,
$K^0$ must vanish in every frame and hence, given the assumed
condition on $K^{\mu}$, so must $K^{\mu}$.}. Under these conditions,
the right hand side of \keqn\  is non-negative and hence so is the
left hand side. This implies that $$ M\ge {\sqrt{3}\over 2}Q\ ,
\eqn\newq $$ where $M=\sqrt{P^{ADM}\cdot P^{ADM}}$ is the ADM mass.
This bound can be saturated
 only if the right hand side of \keqn\ vanishes which, as we have now
seen, requires that $$ \hat\nabla_{\mu}\epsilon =0\qquad K^{\mu}=0\ .
\eqn\vanis $$ Let us define
$V^{\mu}=\bar\epsilon\Gamma^{\mu}\epsilon$. Note that $$
\hat\nabla_{(\mu}V_{\nu )}=\overline{
\left(\hat\nabla_{(\mu}\epsilon\right)}\Gamma_{\nu )}\epsilon
+\bar\epsilon\Gamma_{(\nu}\hat\nabla_{\mu )}\epsilon\ , \eqn\killeqn
$$ so that if $\hat\nabla_{\mu}\epsilon =0$ then $V^{\mu}$ is a
Killing vector field. Note also the identity (for
$V^{\mu}=\bar\epsilon\Gamma^{\mu}\epsilon$), valid in $d=5$, $$
\left(\Gamma\cdot V\right)\epsilon
=sgn\left(\bar\epsilon\epsilon\right)\mid V\mid\epsilon\ , \eqn\newid
$$ which implies, in particular, that $$ V\cdot
V=\left(\bar\epsilon\epsilon\right)^2\ge 0\ . \eqn\vzero $$ Hence,
$V^\mu$ is timelike or null. At infinity \newid\ reduces to $$
\left(\Gamma\cdot V_{\infty}\right)\epsilon_{\infty}
=sgn\left(\bar\epsilon_{\infty}\epsilon_{\infty}\right)\mid
V_{\infty}\mid\epsilon_{\infty}\ . \eqn\newidin $$ We now observe
that the vanishing of the left hand side of \keqn\ requires that
$\epsilon_\infty$ satisfy $$ \left( \Gamma\cdot
P^{ADM}\right)\epsilon_{\infty}= sgn(Q)\mid P^{ADM}\mid
\epsilon_{\infty}\ . \eqn\pnewid $$ By comparison with \newidin\ and
choosing $sgn\left( \bar\epsilon_{\infty}\epsilon_{\infty}\right)
=sgn(Q)$ we deduce that $V_\infty$ is parallel to $P^{ADM}$. For
static spacetimes and matter field configurations for which
$V={\partial\over\partial t}$, equation \newid\ reduces to $$
\Gamma^{\underline 0}\epsilon =\pm\epsilon\ , \eqn\epcond $$ and
$K^{\mu}=0$ is equivalent to $$ T_{\underline 0\underline
0}(mat.)={\sqrt{3}\over 2}\mid J_{\underline 0}(mat.)\mid\ .
\eqn\tcondn $$

In the interior of $\Sigma$, $V$ may become null on an event horizon
(it cannot become spacelike). In this case $\left(\Gamma\cdot
V\right)\epsilon =0$, but this is not consistent with the boundary
condition on $\epsilon$ at event horizons that was used in [\ghhp]
unless $\epsilon$ vanishes at the horizon. However, the condition
$\left(\Gamma\cdot V\right)\epsilon =0$ is only needed to {\it
saturate} the bound \newq . We shall see from our explicit solutions
to follow, that {\it either} there are no horizons when the bound is
saturated {\it or} (in the absence of matter) $\epsilon$ vanishes at
an infinitely distant horizon.

It would be of interest to determine the general configuration
compatible with the constraint $\hat\nabla_{\mu}\epsilon =0$, as has
been done for $d=4$ Einstein/Maxwell [\tod], but we shall be content
here to establish the existence of a simple class of solutions found
via the ansatz $$ \eqalign{ e_0{}^{\underline i}&= e_i{}^{\underline
0}=0\cr e_i{}^{\underline j} &=e^\phi\delta_i^j\qquad \dot\phi=0\cr
\dot e_0{}^{\underline 0}&=0\ .} \eqn\ansatf $$ It follows that $$
\omega_{0}{}_{\underline i\, \underline 0}=-e^{-\phi}\partial_i
e_0{}^{\underline 0}\qquad \omega_i{}_{\underline j\underline k}= -
2\delta_{i[j}\partial_{k]}\phi\ , \eqn\spinf $$ all other components
of the spin connection vanishing. Note that we need not distinguish
between indices $i$ and $\underline i$ on the right-hand sides of
these equations since all tensors and derivative operators become
those of Euclidean $4$-space. Using the condition \epcond\  equation
\kill\ can now be shown to imply $$ e_0{}^{\underline 0}= e^{-2\phi}
\qquad F_{0i}=\mp{\sqrt{3}\over2}\partial_i\big(e^{-2\phi}\big)\ ,
\eqn\mff $$ and $$ \epsilon = e^{-\phi}\epsilon_{\infty}\ , \eqn\eppfl
$$ where $\epsilon_{\infty}$ is a constant spinor which we may
identify, without loss of generality, as the normalised constant
spinor introduced previously. These results imply that the metric has
the form $$ ds^2= e^{-4\phi}dt^2 \, -\, e^{2\phi} d{\bf x}\cdot d{\bf
x}\ , \eqn\metricf $$ where $d{\bf x}\cdot d{\bf x}$ is the Euclidean
$4$-metric, and that the Maxwell potential one-form is $$
A=\pm{\sqrt{3}\over2}e^{-2\phi} dt\ . \eqn\oformf $$ In the following
section we shall determine and solve the conditions on $\phi$ that
are required for this supersymmetric Einstein/Maxwell field
configuration to solve the field equations \einf, and study the
particular case of no matter.

\chapter{d=5 Supersymmetric Black Holes}

Given \metricf\ and \oformf, the left hand sides of the field
equations \einf\ may be computed. The result is $$ \eqalign{
G_{\underline 0\underline 0}-2T_{\underline 0\underline 0}(F)&=
-{3\over2}e^{-4\phi}\square e^{2\phi}\cr \partial_i\big(eF^{i0}\big)
&=\mp{\sqrt{3}\over2}\square e^{2\phi}\ ,} \eqn\bfield $$ all other
components vanishing {\it identically}. Here, $\square$ is the
Laplacian for {\it Euclidean} $4$-space. In the absence of matter we
deduce that $$ \square e^{2\phi}=0\ . \eqn\lape $$ Let us impose the
boundary condition $\phi=0$ at spatial infinity (so
$\epsilon\rightarrow\epsilon_{\infty}$ as required). The general
solution with four-dimensional spherical symmetry is then $$
e^{2\phi} =1+{a^2\over r^2}\ , \eqn\spsoln $$ where $r$ is the radial
distance in the Euclidean $4$-metric and $a$ is a constant. The
$5$-metric \metricf\ is then $$ ds^2 = \left( 1+{a^2\over
r^2}\right)^{-2} dt^2\, -\, \left( 1+{a^2\over r^2}\right) d{\bf
x}\cdot d{\bf x}\ , \eqn\bmetric $$ Rewriting this metric in terms of
a new radial coordinate $\tilde r$ defined by $r^2= \tilde r^2 -a^2$,
it becomes $$ ds^2 = \left( 1-{a^2\over \tilde r^2}\right)^2 dt^2 -
\left( 1-{a^2\over \tilde r^2}\right)^{-2}d\tilde r^2 - \tilde r^2
d\Omega_3^2\ , \eqn\rbmetric $$ where $d\Omega_3^2$ is the round
metric on the unit $3$-sphere. Substitution of this metric into the
expression given earlier for the total mass $M$ yields $$ a^2= {4G_5
M\over 3\pi}\ . \eqn\bmass $$ This metric was found previously in
[\myers]. Here we have shown that it is supersymmetric in the sense
that it admits a Killing spinor.

To analyse the behaviour as $\tilde r\rightarrow a$ we define the new
dimensionless variable $\lambda$ by $$ \tilde
r=a\left(1+{\lambda\over 2}\right)\ . \eqn\dimless $$ Then the metric
becomes $$ ds^2=\left[\lambda^2dt^2-{a^2\over
4\lambda^2}d\lambda^2-a^2d\Omega_3^2\right]\left\{ 1+O(\lambda
)\right\}\ . \eqn\newmetfg $$ The asymptotic form of the metric near
$\lambda =0$ is found by neglecting the $O(\lambda )$ terms in this
expression. The resulting metric is simply that of ${\rm ad}S_2\times
S^3$. This fact implies, incidentally, that ${\rm ad}S_2\times S^3$ is
an allowed compactification of pure $d=5$ supergravity. Thus, as for
the $d=4$ RN black hole [\gibbo], the $d=5$ black hole metric
\bmetric\ interpolates between two vacuum solutions of $d=5$
Einstein/Maxwell theory.

\chapter{Supersymmetric Self-Gravitating Solitons For $d=5$}

It is clear from the result summarised in \bfield\ that any `matter'
source for the Einstein/Maxwell equations that is consistent with the
existence of a Killing spinor must be such that {\it the only
non-vanishing components of the matter stress-tensor and electric
current are $T_{00}(mat.)$ and $J_0(mat.)$} and, furthermore, that $$
T_{\underline 0\underline 0}(mat.)= {\sqrt{3}\over2}\mid
J_{\underline 0}(mat.)\mid\ . \eqn\smat $$ Note that such matter
currents satisfy the condition $K^{\mu} =0$ of Section 2 which we
found there to be necessary for saturation of the energy bound
$M={\sqrt{3}\over 2}\mid Q\mid$. This condition is satisfied by
`extremal charged dust' but here we are interested in matter in the
form of field theory solitons. We now specify the `matter' to be a
Lie-algebra-valued YM potential $B_\mu$ with field-strength
$G_{\mu\nu}$\foot{It should be obvious from the context when
$G_{\mu\nu}$ refers to the field-strength of $B_\mu$ and when it
refers to the Einstein tensor.} $$ G_{\mu\nu}=\partial_\mu B_\nu
-\partial_\nu B_\mu +\left[ B_\mu ,B_\nu\right]\ . \eqn\slop $$ We
shall take the action to have the form $$ S_{matter} = {1\over
g_5^2}\int\!d^5x\left\{ -{1\over 4} e\, \tr
\big(G_{\mu\nu}G^{\mu\nu}\big)
+{c\over4\sqrt{3}}\varepsilon^{\mu\alpha\beta\gamma\delta}
A_\mu\tr\big(G_{\alpha\beta}G_{\gamma\delta}\big)\right\}\ ,
\eqn\actmat $$ for some dimensionless constant $c$\foot{Note that $c$
{\it is not} the speed of light, which has been set to $1$
throughout.}. The YM field will be assumed to have dimensions of
$L^{-1}$, so the $d=5$ YM coupling constant $g_5$ has dimensions of
$M^{-(1/2)}$. From this action, it follows that $$ \eqalign{
T_{\mu\nu}(mat.)&= T_{\mu\nu}(G)\equiv -{1\over g_5^2}\tr\left(
G_{\mu\lambda}G_\nu{}^\lambda -{1\over4}
g_{\mu\nu}G^{\alpha\beta}G_{\alpha\beta}\right)\cr J^\mu(mat.)
&=J^\mu (G)\equiv {c\over 4\sqrt{3}g_5^2}e^{-1}
\varepsilon^{\mu\alpha\beta\gamma\delta}
\tr\big(G_{\alpha\beta}G_{\gamma\delta}\big)\ .} \eqn\fieldmat $$

Consider any YM field configuration for which $G_{i0}=0$ and $$
\sqrt{g^{(4)}}G^{ij}=\pm{1\over2}\varepsilon^{ijkl}G_{kl}\ ,
\eqn\ymmat $$ where $g^{(4)}= \det g_{ij}$. It is not difficult to
show that for such configurations all components of $T_{\mu\nu}(G)$
and $J_\mu(G)$ vanish except $T_{00}(G)$ and $J_0(G)$ and,
furthermore, that $$ T_{\underline 0\underline 0}(G)=
{c\sqrt{3}\over2}\mid J_{\underline 0}(G)\mid\ . \eqn\smatc $$ Hence,
to satisfy \smat\ we must set $$ c=1\ . \eqn\scno $$ This is
precisely the value given by coupling super YM theory to $d=5$
supergravity [\gst ].

It remains to check whether \ymmat\ satisfies the YM field equation.
This is $$ D_\mu\big(eG^{\mu\nu}\big) \, +\, {c\over
2\sqrt{3}}\varepsilon^{\nu\alpha\beta
\rho\sigma}F_{\alpha\beta}G_{\rho\sigma} =0\ , \eqn\ymfield $$ where
$D_\mu$ is the YM covariant derivative. Since $F$ is purely electric
and $G$ is purely `magnetic' the time component of this equation is
trivially satisfied. The space component reads $$ D_i\big(eG^{ij}\big)
-{c\over\sqrt{3}}\varepsilon^{jklm}F_{0k}G_{lm} =0\ . \eqn\ymspace $$
Into this equation we substitute $$ e=e_0{}^{\underline
0}\sqrt{g^{(4)}} = e^{-2\phi}\sqrt{g^{(4)}} \qquad \qquad
F_{0i}=\mp{\sqrt{3}\over2}\partial_i\big(e^{-2\phi}\big)\ , \eqn\eym
$$ to reduce it to $$
e^{-2\phi}\partial_i\left(\sqrt{g^{(4)}}G^{ij}\right)
-\partial_j\big(e^{-2\phi}\big)\left(\sqrt{g^{(4)}}G^{ij} \mp{c\over
2}\varepsilon^{ijkl}G_{kl}\right) =0\ , \eqn\redym $$ which is solved
by any solution of \ymmat, provided again that $c=1$.

Observe that for any conformally flat $4$-metric, \ymmat\ is
equivalent to the {\it Euclidean $4$-space} YM self-duality equation.
$$ G_{ij} = \pm {1\over2}\varepsilon^{ijkl}G_{kl}\ . \eqn\sdym $$ We
have therefore shown that any {\it flat space} YM instantonic-soliton
configuration, together with a metric and Maxwell one-form of the form
\metricf\ and \oformf, is a supersymmetric solution of the
Euler-Lagrange equations of the action $$ \eqalign{ S=&{1\over 4\pi
G_5}\int\! d^5x\left\{ -{1\over4}eR -{1\over4}eF_{\mu\nu}F^{\mu\nu} +
{1\over 6\sqrt{3}}\varepsilon^{\mu\alpha\beta\gamma\delta} A_\mu
F_{\alpha\beta}F_{\gamma\delta}\right\} \cr &+{1\over g_5^2}\int\!
d^5x\left\{ -{1\over4} e\, \tr \big(G_{\mu\nu}G^{\mu\nu}\big)
+{1\over4\sqrt{3}}\varepsilon^{\mu\alpha\beta\gamma\delta}
A_\mu\tr\big(G_{\alpha\beta}G_{\gamma\delta}\big)\right\}\ .}
\eqn\elym $$ Note that this includes multi-soliton solutions.

To fully specify the solution we must determine $\phi$ for a given
soliton source. To do this we return to \einf\ and use the results
\bfield\ and \fieldmat\ to find that $$ \square e^{2\phi}
=-{1\over6}\left({4\pi G_5\over
g_5^2}\right)\Bigm|\varepsilon^{ijkl}\tr\big(G_{ij}G_{kl}\big)\Bigm|\
, \eqn\lapym $$ which now replaces \lape. For any choice of instanton
or multi-instanton localised within a finite `core' region of
$4$-space, the solution of this equation has the asymptotic form $$
e^{2\phi} \sim 1+ \left({16\pi G_5\over
3g_5^2}\right){\mid\nu\mid\over r^2}\ , \eqn\asymym $$ where $\nu$ is
the instanton number $$ \nu ={1\over 16\pi^2}\int\!\!d^4x\,
\varepsilon^{ijkl}\tr\big(G_{ij}G_{kl}\big) \ . \eqn\instym $$
Comparison with the extreme $d=5$ black hole solution of the previous
section yields the result $$ M= \left({2\pi\over g_5}\right)^2
\mid\nu\mid\ . \eqn\bhym $$ For non-zero core radius (or radii in the
multi-soliton case) the resulting spacetime is non-singular and
without event horizons. In the limit of vanishing soliton core radius
the asymptotic solution \asymym\ becomes exact and we recover the
extreme $d=5$ RN solution with  mass $M$. Note that not all values of
the mass are obtained in this way because \bhym\ is a quantisation
condition. Note also that since $\mid Q\mid =(2/\sqrt{3})M$, we also
have the relation $$ \mid Q\mid = {8\pi^2\over\sqrt{3}g_5^2}\,
\mid\nu\mid\  \eqn\qym $$ between the electric charge $Q$ and the
soliton's topological charge $\nu$. Finally we remark that if the
$4$-space is asymptotically {\bb R}${}^3\times S^1$ instead of {\bb
R}${}^4$, solutions of \sdym\ yield (multi) monopole solutions of the
dimensionally reduced $d=4$ theory, but we shall deal directly with
the $d=4$ case in the following sections.

\chapter{An Energy Bound For $d=4$ Einstein/Maxwell Theory With
Arbitrary Dilaton Coupling}

Motivated by consideration of the dimensional reduction of $d=5$
supergravity to $d=4$ we now turn our attention to a $d=4$ action of
the form $$ S={1\over 4\pi G}\int d^4x\, e\left[ {1\over4}R
-{1\over4} e^{2b\sigma} F^2 -{1\over2} (\partial\sigma)^2 \right] \
+\ S_{matter}\ , \eqn\actionw $$ where $\sigma$ is the dilaton field
and $b$ the dilaton coupling constant.  We might also have included
an axion field arising from the fifth component of the $d=5$ Maxwell
potential but since only $A_0$ was non-zero for the $d=5$ solutions
found above, we omit it. The $d=4$ Newton's constant $G$ has
dimensions of $LM^{-1}$. {\it Note also that we now use a `mostly
plus' signature}.

The dimensional reduction and subsequent truncation of $d=5$
supergravity to an action of the form \actionw\ yields a value of
$\sqrt {3}$ for the dilaton coupling constant $b$, and if $b=-1$ the
action \actionw\ is a truncation of $N=4$ supergravity. At least for
these values of $b$, there is an underlying $N=2$ supergravity model.
We believe that \actionw\ is the truncation of some model of $N=2$
supergravity coupled to a scalar multiplet for any value of $b$. All
the results to follow are consistent with this hypothesis, but we
have not checked it explicitly.

The field equations of \actionw\ are $$ \eqalign{ G_{\mu\nu}
-2T_{\mu\nu}(F) -2T_{\mu\nu}(\sigma) &= (8\pi G)T_{\mu\nu}(mat.)\cr
\nabla_\mu\big( e^{2b\sigma}F^{\mu\nu}\big) &= (4\pi G) J^\nu(mat.)\cr
\partial_\mu\big( eg^{\mu\nu}\partial_\nu\sigma\big)
-{g\over2}ee^{2b\sigma} F^{\mu\nu}F_{\mu\nu} &= -(4\pi G){\delta
S_{matter}\over\delta\sigma}\ ,} \eqn\fieldw $$ where $$ \eqalign{
T_{\mu\nu}(F) &= e^{2b\sigma}\left( F_{\mu\lambda}F_\nu{}^\lambda
-{1\over4} g_{\mu\nu}F^2\right)\cr T_{\mu\nu}(\sigma) &=
\left(\partial_\mu\sigma\partial_\nu\sigma
-{1\over2}g_{\mu\nu}(\partial\sigma)^2\right)\ ,} \eqn\engmw $$ and $$
T_{\mu\nu}(mat.) \equiv {1\over e}{\delta S_{matter}\over\delta
g^{\mu\nu}} \qquad J^\mu(mat.) \equiv -{1\over e} {\delta
S_{matter}\over \delta A_\mu} \ . \eqn\varnw $$

As we did for $d=5$, we now introduce the Nester-like tensor $\hat
E^{\mu\nu}$ defined as before in terms of a modified covariant
derivative $\hat \nabla_\mu$ acting on a complex spinor $\epsilon$; $$
\hat E^{\mu\nu} ={1\over2}\bar\epsilon
\Gamma^{\mu\nu\rho}\hat\nabla_\rho\epsilon \ + c.c.\ . \eqn\nesterw $$
For $d=4$, and including the dilaton field, this modified covariant
derivative is $$ \hat \nabla_\mu\epsilon = \nabla_\mu\epsilon +
{i\over 4\sqrt{1+b^2}} e^{b\sigma}\Gamma^{\alpha\beta}
\Gamma_\mu\epsilon\, F_{\alpha\beta}\ . \eqn\covw $$ It will also be
necessary to define the quantity $$ \delta\lambda \equiv
{1\over\sqrt{2}}\left[ \Gamma^\mu\epsilon\, \partial_\mu\sigma +
{ib\over 4\sqrt{1+b^2}}e^{b\sigma}\Gamma^{\alpha\beta}\epsilon\,
F_{\alpha\beta}\right]\ . \eqn\dellam $$ The specific factors that
appear in these definitions are motivated {\it a posteriori} as those
that are required to derive the energy bound to follow, but they also
have an {\it a priori} motivation as the supersymmetry transformation
laws of the gravitino and dilatino fields in the associated
supergravity model.

As before, let us choose $\Sigma$ to be a spacelike hypersurface with
hypersurface element $dS_{\mu}$ and boundary $\partial\Sigma$ at
spatial infinity. We can write $\hat E^{\mu\nu}$ as $$ \hat
E^{\mu\nu}={1\over
2}\left\{\bar\epsilon\Gamma^{\mu\nu\rho}\nabla_{\rho}\epsilon
-{i\over {\sqrt{1+b^2}}}\bar\epsilon\left( F^{\mu\nu}+{1\over
2}\Gamma^{\mu\nu\alpha\beta}F_{\alpha\beta}\right)\epsilon
e^{b\sigma}\right\} + c.c.\ . \eqn\mengw $$ Assuming that $\epsilon$
is asymptotic to a {\it constant spinor} $\epsilon_{\infty}$ and that
$\sigma\sim 0$ at spatial infinity, we deduce (using the fact that
$e\Gamma^{\mu\nu\alpha\beta}=\varepsilon^{\mu\nu\alpha\beta}\gamma_5$
where $\gamma_5 =\Gamma^{\underline 0}\cdots\Gamma^{\underline 3}$)
that $$ \eqalign{ \int_\Sigma \!\! dS_\mu\, e\nabla_\nu\hat
E^{\mu\nu} &=\int_{\partial\Sigma}\!\! dS_{\mu\nu}\left[ \fourth
e\bar\epsilon_{\infty}\Gamma^{\mu\nu\rho}\nabla_{\rho}\epsilon_{\infty}
+c.c.\right]\cr   &- {i\over{2\sqrt{1+b^2}}}
\int_{\partial\Sigma}\!\!dS_{\mu\nu}\,  \left[
e\bar\epsilon_{\infty}\epsilon_{\infty} F^{\mu\nu} +
\bar\epsilon_{\infty}\gamma_5\epsilon_{\infty}\tilde
F^{\mu\nu}\right]\cr &=
\left(\bar\epsilon_{\infty}\Gamma^{\mu}\epsilon_{\infty}\right)
P_{\mu}^{ADM} -{i\over\sqrt{1+b^2}}\bar\epsilon_{\infty}\left(
Q+\gamma_5 P\right)\epsilon_{\infty} \ ,} \eqn\hypengw $$ where
$P^{ADM}_{\mu}$ is the ADM $4$-momentum, $\tilde F^{\mu\nu}={1\over
2}\varepsilon^{\mu\nu\alpha\beta}F_{\alpha\beta}$ and  $$
Q=\half\int_{\partial\Sigma}\!\!dS_{\mu\nu}\left( eF^{\mu\nu}\right)
\qquad P=\half\int_{\partial\Sigma}\!\!dS_{\mu\nu}\tilde F^{\mu\nu}\ ,
\eqn\qcharg $$ are the total electric and magnetic charges
respectively.

A lengthy calculation yields the result $$ \nabla_{\nu}\hat
E^{\mu\nu}=\left\{
-\half\overline{\left(\hat\nabla_{\nu}\epsilon\right)}\Gamma^{\mu\nu\rho}
\hat\nabla_{\rho}\epsilon
-\half\overline{\delta\lambda}\Gamma^{\mu}\delta\lambda +c.c.\right\}
-\left( 4\pi G\right)K^{\mu}\ , \eqn\newlong $$ where $$
K^{\mu}=\bar\epsilon\Gamma^{\nu}\epsilon T_{\nu}{}^{\mu}(mat.)+{i
e^{-b\sigma}\over\sqrt{1+b^2}}\bar\epsilon\epsilon J^{\mu}(mat.)\ .
\eqn\klong $$ Now upon using equations \hypengw\ and \newlong\ we
deduce that $$ \eqalign{
\left(\bar\epsilon_{\infty}\Gamma^{\mu}\epsilon_{\infty}\right)P^{ADM}_{\mu}
- {i\over{\sqrt{1+b^2}}}\bar\epsilon_{\infty}&\left( Q+\gamma_5
P\right)\epsilon_{\infty} = \int_{\Sigma}\!\! dS_{\mu}e \Biggl[
-\half\overline{\left(\hat\nabla_{\nu}\epsilon\right)}\Gamma^{\mu\nu\rho}
\hat\nabla_{\rho}\epsilon \cr
&-\half\overline{\delta\lambda}\Gamma^{\mu}\delta\lambda +c.c.\Biggr]
-\left( 4\pi G\right)\int_{\Sigma}\!\! dS_{\mu}eK^{\mu}\ ,} \eqn\adml
$$ The first term on the right hand side of \adml\ is non-negative
for spinors $\epsilon$ satisfying the (modified) Witten condition $$
n\cdot\hat\nabla\epsilon =0\ , \eqn\nmodwitt $$ where $n$ is a
$4$-vector normal to $\Sigma$. The last
 term on the right hand side of \adml\ is non-negative if $K^{\mu}$ is
future-directed timelike for all $\epsilon$, and henceforth we shall
assume this to be the case.

Under these conditions the right hand side of \adml\ is non-negative,
and vanishes if and only if $$ \hat\nabla_{\mu}\epsilon =0\qquad
\delta\lambda =0\qquad K^{\mu}=0\ . \eqn\lotscond $$ Hence the left
hand side of \adml\ is non-negative. This implies that $$
M\ge{1\over\sqrt{1+b^2}}\sqrt{Q^2+P^2}\ , \eqn\mqpbd $$ and the bound
is saturated when the conditions \lotscond\ hold. Saturation of the
bound implies that the left hand side of \adml\ vanishes, so that
$\epsilon_\infty$ satisfies $$
\left(\Gamma^{\mu}P_{\mu}^{ADM}\right)\epsilon_{\infty}=
{i\over\sqrt{1+b^2}}\left( Q+\gamma_5 P\right)\epsilon_{\infty}\ .
\eqn\admcond $$ As in the $d=5$ case, we define $$
V^{\mu}=\bar\epsilon\Gamma^{\mu}\epsilon\ , \eqn\vdef $$ which is a
Killing vector field when $\hat\nabla_{\mu}\epsilon =0$. Note the
identity, valid in $d=4$, $$ \left(\Gamma\cdot V\right)\epsilon =
\left[ \bar\epsilon\epsilon +\left(\bar\epsilon\gamma_5\epsilon\right)
\gamma_5\right]\epsilon\ , \eqn\nepcond $$ which implies that $$
V\cdot V=\left(\bar\epsilon\epsilon\right)^2+
\left(\bar\epsilon\gamma_5\epsilon\right)^2\ge 0\ . \eqn\vnep $$
Hence $V$ is timelike or null. At infinity \nepcond\ reduces to
\admcond\ with $V\propto P^{ADM}$ and $$ {P\over
Q}={\bar\epsilon_{\infty}\gamma_5\epsilon_{\infty}\over
\bar\epsilon_{\infty}\epsilon_{\infty}}\ . \eqn\pqepcond $$

We shall be interested in static spacetimes admitting a Killing
spinor and with $P=0$. In this case we can choose
$\bar\epsilon\gamma_5\epsilon =0$. Then equation \nepcond\ reduces
to\foot{Observe that $\left(i\Gamma^{\underline 0}\right)^2=1$ for
the `mostly plus' metric signature that we use for $d=4$.}  $$
i\Gamma^{\underline 0}\epsilon =\pm\epsilon\ , \eqn\uiyo $$ where the
timelike Killing vector field $V$ is given by
$V={\partial\over\partial t}$. In this case the condition $K^{\mu}=0$
reduces to $$ T_{\underline 0\underline
0}(mat.)={e^{-b\sigma}\over\sqrt{1+b^2}}\mid J_{\underline
0}(mat.)\mid\ . \eqn\fourmat $$ Note that \uiyo\ is not compatible
with the boundary condition on $\epsilon$ at event horizons, but
again \uiyo\ is only required for {\it saturation} of the bound
\mqpbd . As for $d=5$, spacetimes that saturate the bound either have
no horizons or horizons at an infinite distance at which $\epsilon$
vanishes.

We shall again seek solutions that can be found via the ansatz
\ansatf. We should note that the change of signature from the five
dimensional case means that the components of the spin connection now
become $$ \omega_{0\underline i\underline
0}=e^{-\phi}\partial_ie_0{}^{\underline 0}\qquad \omega_{i\underline
j\underline k}=+2\delta_{i[j}\partial_{k]}\phi\ . \eqn\spinw $$ This
now leads to a metric of the form $$ ds^2 = - e^{-2\phi} dt^2 +
e^{2\phi}d{\bf x}\cdot d{\bf x}\ , \eqn\metricw $$ where $d{\bf
x}\cdot d{\bf x}$ is the Euclidean $3$-metric. It leads also to the
Maxwell one-form $$ A = \pm {1\over \sqrt{1+b^2}} e^{-(1+b^2)\phi}
dt\ , \eqn\oformw $$ and to the dilaton $$ \sigma =b\phi\ . \eqn\dilw
$$ The solution of $\hat\nabla_{\mu}\epsilon =0$, $\delta\lambda =0$
is then found to be $$ \epsilon =
e^{-{1\over2}\phi}\epsilon_{\infty}\ , \eqn\solutionw $$ where
$\epsilon_{\infty}$ is the constant spinor introduced previously,
since we shall impose the boundary condition that $\phi\sim 0$ as
spatial infinity is approached.

\chapter{Supersymmetric Dilaton Black Holes For d=4}

Given the results just quoted, the Einstein, Maxwell and dilaton
field equations of the action \actionw\ become $$ \eqalign{ (4\pi
G)T_{{\underline 0}{\underline 0}}(mat.) &= -{1\over 1+b^2}
e^{-(3+b^2)\phi} \nabla^2 \left[ e^{(1+b^2)\phi}\right]\cr (4\pi
G)e^{-b\sigma}J_{\underline 0}(mat.) &= \mp {1\over
\sqrt{1+b^2}}e^{-(3+b^2)\phi}\nabla^2 \left[ e^{(1+b^2)\phi}\right]\cr
(4\pi G){\delta S_{matter}\over\delta \sigma}&= -{b\over 1+b^2}
e^{-(1+b^2)\phi}\nabla^2 \left[ e^{(1+b^2)\phi}\right]\ ,}
\eqn\sfieldw $$ where $\nabla^2$ is the Laplacian of
three-dimensional Euclidean space. In the absence of matter all three
equations reduce to $$ \nabla^2 \left[ e^{(1+b^2)\phi}\right]=0\ .
\eqn\lapew $$ The spherically symmetric solution corresponding to the
boundary condition that $\phi$ vanish at spatial infinity is $$
e^{(1+b^2)\phi}= 1+ {\mu \over r}\ , \eqn\expsolw $$ for constant
$\mu$. Rewriting the metric in terms of the new radial coordinate
$\tilde r$, now defined by $r= \tilde r -\mu$, we find that $$ ds^2 =
-\left(1-{\mu\over \tilde r}\right)^{2\over 1+b^2}\, dt^2 +
\left(1-{\mu\over \tilde r}\right)^{-{2\over 1+b^2}}\, d{\tilde r}^2
+ \left(1-{\mu\over \tilde r}\right)^{2b^2\over 1+b^2}{\tilde r}^2
d\Omega_2^2\ , \eqn\cmetricw $$ where $d\Omega_2^2$ is the `round'
metric on $S^2$ and the dilaton is given by $$ e^{\sigma}=\left(
1-{\mu\over\tilde r}\right)^{-{b\over 1+b^2}}\ . \eqn\newdilato $$
Substitution of \cmetricw\ into the Witten-Nester expression for the
total mass $M$ yields the relation $$ \mu = (1+b^2)GM\ , \eqn\massrw
$$ between $M$ and the constant $\mu$. This extreme `dilatonic' black
hole solution has been found previously [\shiraishi]. Here we have
shown that it is supersymmetric.

The metric \cmetricw\ is singular at $\tilde r=\mu$, but the rescaled
metric $e^{2b\sigma}ds^2$ is non-singular. Using equations \cmetricw\
and \newdilato\ we find the rescaled metric to be  $$
e^{2b\sigma}ds^2=-\left(1-{\mu\over \tilde r}\right)^{2(1-b^2)\over
(1+b^2)}\, dt^2 +\left(1-{\mu\over \tilde r}\right)^{-2}\, d{\tilde
r}^2+{\tilde r}^2 d\Omega_2^2\ . \eqn\rescled $$ To analyse the
behaviour as $\tilde r\rightarrow\mu$ we define the new
(dimensionless) variable $\lambda$ by $$ \tilde
r=\mu\left(1+\lambda\right)\ . \eqn\newresd $$ Then the rescaled
metric becomes $$ e^{2b\sigma}ds^2=\left[ -\lambda^{{2( 1-b^2)\over (
1+b^2)}}\,
dt^2+\left({\mu\over\lambda}d\lambda\right)^2+\mu^2d\Omega_2^2
\right]\left\{ 1+O\left(\lambda\right)\right\}\ . \eqn\relamd $$ The
asymptotic form of the metric near $\lambda  =0$ is found by
neglecting the $ O\left(\lambda\right)$ terms in this expression.
Thus, introducing the new variable $\rho =\mu\ln\lambda$, we see that
the rescaled metric has the asymptotic form  $$ e^{2b\sigma}ds^2\sim
-e^{-{2(1-b^2)\over (1+b^2)}{\rho\over\mu}} \, dt^2
+d\rho^2+\mu^2d\Omega_2^2\ , \eqn\finres $$ and the dilaton has the
asymptotic form $$ \sigma\sim {-b\over (1+b^2)} {\rho\over\mu}\ .
\eqn\redila $$ If $b\neq\pm 1$ the resulting metric is simply that of
$adS_2\times S^2$, whilst if $b=\pm 1$ it is a metric for ${\cal
M}^2\times S^2$. Hence dilatonic extreme black holes interpolate
between four-dimensional Minkowski space and a compactified vacuum
spacetimes with a linear dilaton, at least for one choice of
`conformal gauge'. This behaviour is similar to that of the extreme
fivebrane solution of $d=10$ supergravity in string-theory conformal
gauge [\gibtow].

\chapter{Supersymmetric Self-Gravitating Monopoles With A Dilaton}

We now return to the dilatonic action of $(6.1)$ and take the matter
action to be $$ \eqalign{ S_{matter} =& {1\over g^2}\int\!\! d^4
x\bigg\{ -{1\over4} e^{{(1-b^2)\over b}\sigma} e\, \tr
(G_{\mu\nu}G^{\mu\nu}) -{1\over2} e^{-{(1+b^2)\over b}\sigma}e\, \tr
(D_\mu\Phi D^\mu\Phi)\cr &+
{c\sqrt{1+b^2}\over4}\varepsilon^{\mu\nu\alpha\beta}F_{\mu\nu}\tr(\Phi
G_{\alpha\beta}) \bigg\}\ ,} \eqn\actmatw $$ where $c$ is a constant
(again, not to be confused with the speed of light which is set equal
to unity) soon to be determined. The particular coupling of the
dilaton to the YM/Higgs fields in \actmatw\ are those required for
consistency of the YM/Higgs equations with the Einstein/Maxwell
/dilaton equations and the assumption that the metric, Maxwell
one-form and dilaton have the form found in Section $6$. From
\actmatw\ and the definitions \varnw\ we have that  $$
\eqalign{T_{\mu\nu}(mat.)&={1\over g^2}e^{{(1-b^2)\over b}\sigma}\tr
\left( G_{\mu\lambda}G_{\nu}{}^{\lambda}-{1\over 4}g_{\mu\nu}
G_{\alpha\beta}G^{\alpha\beta}\right)\cr &+ {1\over
g^2}e^{-{(1+b^2)\over b}\sigma}\tr\left( D_{\mu}\Phi D_{\nu}\Phi
-{1\over 2}g_{\mu\nu}D_{\alpha}\Phi D^{\alpha}\Phi \right)\cr
J^{\nu}(mat.)&={ce^{-1}\over
2g^2}\sqrt{1+b^2}\varepsilon^{\nu\mu\rho\sigma}\tr\left( D_{\mu}\Phi
G_{\rho\sigma}\right)\ . } \eqn\matfieldw $$

Observe that when $b=\sqrt{3}$, which is what would be found by
reduction (and truncation) to $d=4$ of $d=5$ supergravity, then the
dilaton coupling to the YM field is $1/\sqrt{3}$, which is also what
is found on dimensional reduction from $d=5$. The choice of
Lagrangian \actmatw\ is therefore consistent with what we would find
on reduction of supergravity/YM from $d=5$ but we are now allowing
for arbitrary {\it non-zero} $b$. Once the coupling of the dilaton to
the YM field is chosen its coupling to the Higgs field can be fixed,
and has been so fixed in \actmatw, by requiring invariance of the
action under the rigid scaling $\sigma\rightarrow \sigma + const.$,
which is achieved by assigning scaling weights to all fields and then
choosing powers of $\sigma$ to compensate any lack of scale
invariance. What makes this non-trivial is the presence of the
non-minimal terms in the action involving epsilon tensors for which
the scale weight must add to zero. This symmetry is not necessary for
consistency of the theory, of course, but it would be required by
supersymmetry.

Using the relation $\sigma = b\phi$, from \dilw , the equations of
motion for $B$ and $\Phi$ can be written as $$ D_\mu\left[ e e^{(
1-b^2)\phi}G^{\mu\nu}\right] -
 e^{-(1+b^2)\phi}e\left[\Phi,D^\nu \Phi\right]
-{c\over2}\sqrt{1+b^2}\, \varepsilon^{\alpha\beta\mu\nu}
F_{\alpha\beta}D_\mu\Phi =0\ , \eqn\emotw $$ and $$ D_\mu\left[ e
e^{-(1+b^2)\phi}D^\mu\Phi\right] + {c\sqrt{1+b^2}\over
4}\varepsilon^{\mu\nu\alpha\beta}F_{\alpha\beta}G_{\mu\nu} =0\ .
\eqn\demotw $$ Using \metricw\ and \oformw, and assuming that
$G_{0i}$ and $D_0\Phi$ vanish, these equations  reduce to the {\it
Euclidean $3$-space} equations $$ D_jG_{ji} -\left[\Phi,
D_i\Phi\right] -(1+b^2)\partial_j \phi\left[ G_{ji} \mp c\,
\varepsilon^{ijk} D_k\Phi\right] =0\ , \eqn\euclw $$ and $$ D_i\left(
D_i\Phi\right) -(1+b^2)\partial_i \phi\left[ D_i\Phi \pm
{c\over2}\varepsilon^{ijk} G_{jk}\right] =0\ . \eqn\deuclw $$ {\it
Provided that} $c^2=1$, these equations are solved by any YM-Higgs
configuration that solves the {\it flat space} Bogomol'nyi equations
$$ G_{ij} =\mp c\varepsilon^{ijk}D_k\Phi\ . \eqn\fbogow $$ Here we
see the principal difference with the results of the section 2;
unlike the corresponding equation of that section, equation \fbogow\
is {\it independent} of $\phi$. Its solutions are well known and
include multi-monopole configurations. For example, the one-monopole
solution for YM group $SO(3)$ is $$ \Phi^a=n^a\left[{1\over r}-{\cosh
r\over \sinh r}\right]\qquad\qquad
B_i^a=\varepsilon^{iab}n^b\left[{1\over \sinh r}-{1\over r}\right]\ ,
\eqn\boggosol $$ where $a,b$ are $SO(3)$-vector indices and
$n^a={r^a\over r}$.

As a further illustration of the difference between the dilatonic and
non-dilatonic cases, we may choose $c=-1$ and use the specific form
of the metric ($6.25$) and dilaton ($6.27$) to rewrite \fbogow\ as  $$
e^{{\sigma\over b}}\, \sqrt{g^{{}_{(3)}}} g^{il}g^{jm} G_{lm} = \pm
\varepsilon^{ijk} D_k\Phi\ , \eqn\bogocov $$  which may be compared
with \cbogo . Observe that \bogocov\ has no $b\rightarrow 0$ limit,
so that \cbogo\ cannot be obtained as a limit of \bogocov .

We now turn to the Einstein, Maxwell and dilaton equations. Firstly,
we note that, under the same conditions used to derive \euclw, the
only non-zero components of the matter stress-tensor and electric
current are $$ \eqalign{ T_{{\underline 0}{\underline 0}}(mat.) &=
{1\over g^2}e^{(3+b^2)\phi} \tr (D\Phi\cdot D\Phi)\cr
e^{-b\sigma}J_{\underline 0}(mat.) &= \mp {c\over
g^2}\sqrt{1+b^2}e^{-(3+b^2) \phi}\, \tr (D\Phi\cdot D\Phi)\ ,}
\eqn\stressw $$ where $D\Phi\cdot D\Phi$ indicates the {\it
Euclidean} $3$-vector scalar product. Recall that consistency
requires that $$ T_{{\underline 0}{\underline 0}}(mat.) =
{e^{-b\sigma}\over \sqrt{1+b^2}} \mid J_{\underline 0}(mat.)\mid\ ,
\eqn\consisw $$ which we now see is true {\it provided again that}
$c^2=1$. The Einstein and Maxwell equations may now be seen to be
equivalent to the {\it Euclidean $3$-space} Poisson equation $$
-\nabla^2\left[ e^{\left( 1+b^2\right)\phi}\right] = \left({4\pi
G\over g^2}\right)\left( 1+b^2\right)\tr\left( D\Phi\cdot
D\Phi\right)\ . \eqn\csdq $$ This is also the dilaton equation; the
choice of dilaton coupling to the YM field in \actmatw\ was chosen to
make this happen.

Using the Bogomol'nyi equation, \csdq\ can be rewritten as $$
\nabla^2\left[ e^{(1+b^2)\phi}\right] = -\eta\left({4\pi G\over
g^2}\right)(1+b^2)\mid\partial_i B_i^{mat}\mid\ , \eqn\poissonw $$
where the `matter' magnetic field $B^{mat}_i$ is again given by $$
B^{mat}_i ={1\over2\eta}\varepsilon^{ijk} \tr (\Phi G_{jk})\ .
\eqn\bpoisson $$ The existence of a unique non-singular solution for
$\phi$ to equation \poissonw\ that vanishes at infinity is
guaranteed. We have not been able to find this solution analytically,
but its asymptotic form is $$ e^{(1+b^2)\phi} \sim 1
+(1+b^2)\alpha{\mid P^{mat}\mid\over{\eta r}}\ , \eqn\asypoisson $$
where $P^{mat}$ is the again the (dimensionless) total `matter'
magnetic charge determined by the flux of ${\bf B}^{mat}$ through the
sphere at spatial infinity and $\alpha$ is the dimensionless constant
$$ \alpha ={G\eta^2\over g^2}\ . \eqn\newcst $$  Note that $\phi$ is
non-singular, and hence there are no event horizons, for {\it any}
value of $\alpha$. By comparison with the corresponding formula
\expsolw\ for a black hole, and using \massrw , we see that $$ M =
\left({\eta\over g^2}\right)\mid P^{mat}\mid\ . \eqn\anonmass $$
Since the self-gravitating monopole solution we have found is
supersymmetric and saturates the bound \mqpbd\ with $P=0$, we find
the following relation between the electric charge and the `matter'
magnetic charge: $$ \mid Q\mid=\sqrt{(1+b^2)}\left({\eta\over
g^2}\right)\mid P^{mat}\mid\ . \eqn\anonch $$ The bound on the total
energy implied by supersymmetry may therefore be expressed either in
terms of $Q$ or in terms of $P^{mat}$.

As in section 2, we conclude with the alternative derivation of the
energy bound that is closer in spirit to Bogomol'nyi's original
argument. We insert the ansatz \metricw\ and \oformw\ into the action
\actmatw. From the corresponding Hamiltonian the total energy is then
found to be $$ \eqalign{ E= {1\over g^2}\int &d^3x\bigg\{ {1\over4}
e^{-(1+b^2)\phi} \tr\big( G_{ij}\mp \varepsilon_{ijk}D_k\Phi\big)^2\cr
&+{1\over2}e^{(3-b^2)\phi}\big[\tr (G_{0i} G_{0i}) + \tr (D_0\Phi
D_0\Phi)\big] \pm\eta \partial_i\left[ e^{-(1+b^2)\phi}
B_i^{mat}\right] \bigg\}.} \eqn\ham $$ Assuming that ${\bf
B}^{mat}\sim 1/r$ and that $\phi\rightarrow 0$ as $r\rightarrow
\infty$, we immediately derive the bound $$ E\ge {\eta\over g^2}\mid
P^{mat}\mid\ , \eqn\hbound $$ which is saturated when $$ G_{0i}=0
\qquad D_0\Phi=0\qquad G_{ij}=\pm \varepsilon_{ijk}D_k\Phi\ ,
\eqn\eqsw $$ in agreement with \fbogow . Thus, a solution of \eqsw\
has mass $M$ given by \anonmass.

\chapter{Conclusions}

Minkowski space BPS monopoles saturate a Bogomol'nyi bound on their
mass in terms of their magnetic charge and are therefore stable. \rn\
black holes are stable for a similar reason; they saturate a
gravitational analogue of the Bogomol'nyi bound on their ADM mass in
terms of a combination of electric and magnetic charges. These two
examples of stable `solitons' can be viewed as extreme cases of the
more general situation of a self-gravitating BPS monopole. The
strength of the gravitational field relative to the YM/Higgs fields
is measured by the dimensionless constant $\alpha={G\eta^2\over
g^2}$. If $\alpha\ll 1$, gravitational effects can be ignored and we
have, effectively, a flat space BPS monopole. If $\alpha\gg 1$ one
might expect the monopole core to be hidden behind a horizon, with
only the long range $U(1)$ subgroup of the YM group in evidence, in
which case we have, effectively, a magnetic \rn\ black hole. In the
latter case one expects an {\it approximate} bound on the energy in
terms of the magnetic charge, but this must fail when $\alpha\sim 1$
because the replacement of the YM field by its long range $U(1)$
component is then no longer justifiable. Since the mass of the black
hole is no longer bounded by the magnetic charge one might expect it
to exhibit an instability. Such an instability has been demonstrated
by Lee, Nair and Weinberg [\lnw].

We have shown that the inclusion of certain non-minimal couplings of
the YM/Higgs fields to an additional $U(1)$ `Maxwell' field implies
that the mass of any field configuration {\it is} bounded by the YM
monopole charge. Configurations that saturate this bound are
necessarily stable. This is true whether or not there is a dilaton
field present, but if there is one its coupling to the YM/Higgs
fields cannot be arbitrarily chosen. Accumulated experience suggests
that these non-minimal couplings and the required dilaton couplings
(when applicable) are those which would be required for a coupling of
$N=2$ supergravity to $N=2$ super YM theory. We have not proven this
here but if this proposition is accepted we see that, as expected,
instabilities of the type found in [\lnw] cannot occur in
supersymmetric theories.

Our results for supersymmetric self-gravitating monopoles in the
presence of a dilaton show that there is no event horizon whatever
the monopole core radius, so one cannot `hide' a monopole inside a
black hole. In the absence of a dilaton, the situation is less clear.
It may be that in this case one can `hide' a monopole inside a black
hole. In any event, any black hole/monopole configuration saturating
the gravitational version of the Bogomol'nyi bound is necessarily
stable.

A curious fact that emerges from the analysis of this paper is the
qualitative difference in supersymmetric field configurations
occasioned by the presence of a dilaton. With a dilaton the full
gravitationally-corrected YM/Higgs equations are solved by any
solution of the {\it flat-space} Bogomol'nyi equations. We mentioned
in the introduction that this result can be understood, via
dimensional reduction, as being a consequence of the similar result
for $d=5$, at least for a special value of the dilaton coupling
constant $b$. It seems likely that this observation could be extended
to all non-zero values of $b$ by inclusion of a dilaton in $d=5$.

One aim of this work was to understand in the gravitational context
the much studied Bogomol'nyi bounds saturated by Minkowski space
soliton solutions. We believe that we have achieved this fully in
$d=5$ but there remain questions in $d=4$. In particular, we have not
yet found solutions representing self-gravitating dyons. It will
probably be important to fill this gap before attempting to
investigate the nature of the metric on the moduli space of multi
self-gravitating BPS monopoles because Minkowski space results
suggest that dyons could be produced by scattering monopoles. However
this much seems to be rather plausible: as a manifold the moduli
space must be identical to the flat space moduli space, but the
metric may differ. For arbitrary monopole number, $N=2$ supersymmetry
strongly suggests that it is K\" ahler. For two monopoles the
rotational symmetries of flat {\bb R}${}^3$ should give rise to a
diagonal Bianchi IX metric.

\vskip 1cm

\centerline{\bf Acknowledgements}

One of the authors, L.A.J.L., would like to thank E.A.F. Plessiet for
useful and informative discussions.  D.K., P.K.T. and J.T. thank the
Institute for Theoretical Physics for its hospitality and support
under grant NSF-PHY89-04035.  JT is supported in part by grant
NSF-THY-8714-684-A01.

\appendix

In order to show that spherically symmetric monopoles exist for the
system described in Section 2, we have to solve three coupled
equations.  We will show that these equations have solutions, at
least in the limit of small gravitational coupling\foot{Similar
arguments are made in [\perry,\kt] for the existence of gravitating
monopoles solutions without the extra interaction with the Maxwell
field considered here.}. Make the ansatze for the fields (appropriate
to isotropic coordinates)  $$ \Phi^a=\hat{r}^a s(r),\qquad
B_i^a=\epsilon_{iab}{\hat{r}^b\over r}e^{-\phi} \left (
-1+gv(r)\right)\ . \eqn\ansatze $$ The \bogo\ equation \cbogo\ is
then given by $$ \psi \equiv v^{\prime}(r)-gv(r)s(r)e^\phi=0 ,\qquad
\eta\equiv s^{\prime}(r)  -{1\over r^2}\left (- {1\over g}
+gv^2\right ) e^{-\phi}=0\ . \eqn\abogo $$ This can be used to put
the constraint equation in the form $$ \partial_r\left( r^2\partial_r
e^\phi\right) = -{4\pi G\over g^2} \partial_r\left( s( -{1\over g}
+gv^2)\right)\ .  \eqn\integrable $$ Which can, after using equations
\abogo , be integrated to give $$ e^\phi =e^{\phi_0} e^{-2\pi Gs^2
(r)/g^2}\ . \eqn\solution $$ Since $s\rightarrow \eta$ as
$r\rightarrow\infty$, we find that $e^\phi\rightarrow \left(
1+{M\over r}\right)$, where $M={4\pi \eta\over g^2}$ is the ADM mass
of the spacetime.

We now want to argue for the existence of solutions to equations
\abogo\ for $\alpha={G\eta^2\over g^2}$ sufficiently small.
Substitute the expression \solution\ for $e^\phi$ into \abogo . Then
the left hand sides are functionals $\psi$, $\eta$ of $s(r)$, $v(r)$
and the parameter $\alpha$.  Let $\cF(s,v;\alpha )=(\psi,\eta)$. We
seek solutions $s(\alpha )$, $v(\alpha )$ such that $\cF(s,v;\alpha
)=0$. We know that $\cF (\bar{s},\bar{v};0)=0$, where $\bar{s}$ and
$\bar{v}$ are the known flat space solutions. Hence, by the implicit
function theorem, it suffices to show that (a) $\left.{\partial
\cF\over\partial \alpha}\right| _{\alpha=0} $ is continuous and that
(b) there are no zero modes of $$ \cD\cF\equiv
\left.\left[{\partial\cF\over\partial s}\delta s +
{\partial\cF\over\partial v}\delta v\right] \right| _{\alpha=0} ,
\eqn\zeromode $$ or of $\cD\cF ^*$.  Condition (a) is easily checked
directly, using the behavior of $\bar{s}$ and $\bar{v}$.  Condition
(b) follows from the stability of the flat space solution to small
perturbations.

\refout

\end